\documentclass[aps,pre,longbibliography,showpacs,amsmath,amssymb]{revtex4-1}
\usepackage{amsmath,amssymb,latexsym,epsfig,graphicx,epsf,bm}
\usepackage{graphicx}
\usepackage{float}
\usepackage{inputenc}
\inputencoding{ansinew}
\usepackage[T1]{fontenc}
\usepackage{graphicx}
\usepackage{palatino} 
\usepackage{mathpazo} 
\usepackage{bm} 
\usepackage{color}

\newcommand{\be}{\begin{equation}}
\newcommand{\ee}{\end{equation}}
\newcommand{\fig}[1]{Fig.~\ref{#1}}
\newcommand{\Fig}[1]{Figure~\ref{#1}}

\newcommand{\eq}[1]{Eq.~(\ref{#1})} 
\newcommand{\Eq}[1]{Equation~(\ref{#1})}

\begin{document}
	\title{Rheological model for the alpha relaxation of glass-forming liquids and its comparison to data for DC704 and DC705}
	\date{\today}
	\author{Tina Hecksher}\email{tihe@ruc.dk}
	\affiliation{\textit{Glass and Time}, IMFUFA, Department of Science and Environment, Roskilde University, P.O. Box 260, DK-4000 Roskilde, Denmark}
	\author{Niels Boye Olsen}
	\affiliation{\textit{Glass and Time}, IMFUFA, Department of Science and Environment, Roskilde University, P.O. Box 260, DK-4000 Roskilde, Denmark}	
	\author{Jeppe C. Dyre}\email{dyre@ruc.dk}
	\affiliation{\textit{Glass and Time}, IMFUFA, Department of Science and Environment, Roskilde University, P.O. Box 260, DK-4000 Roskilde, Denmark}

\begin{abstract}
Dynamic shear-modulus data are presented for the two silicone oils DC704 and DC705 for frequencies between 1 mHz and 10 kHz at temperatures covering more than five decades of relaxation-time variation. The data are fitted to the alpha part of a phenomenological model previously shown to describe well the dynamic shear modulus of squalane, which has a large beta process [Hecksher \textit{et al.}, J. Chem. Phys. \textbf{146}, 154504 (2017)]; that model is characterized by additivity of the alpha and beta shear compliance and by a high-frequency decay of the alpha process in proportion to $\omega^{-1/2}$ in which $\omega$ is the angular frequency. The fits of the alpha part of this model to the DC704 and DC705 data are compared to fits by a Havriliak-Negami type model, the Barlow-Erginsav-Lamb model, and a Cole-Davidson type model. At all temperatures the best fit is obtained by the alpha part of the squalane model. This strengthens the conjecture that so-called $\sqrt{t}$-relaxation, leading to high-frequency decays proportional to $\omega^{-1/2}$, is a general characteristic of the alpha relaxation of supercooled liquids [Dyre, Phys. Rev. E {\bf 74}, 021502 (2006); Nielsen \textit{et al.}, J. Chem. Phys. \textbf{130}, 154508 (2009); Pabst \textit{et al.}, J. Phys. Chem. Lett. \textbf{12}, 3685 (2021)].	
\end{abstract}		
\date{\today}
\maketitle

\section{Introduction}

Dynamic shear properties are central for the description of glass-forming liquids \cite{BEL,har76,chr95,dee99,jak05,dyr06}. Historically, many different expressions have been fitted to the dynamic shear modulus $G(\omega)=G'(\omega)+iG''(\omega)$ in which $\omega$ is the angular frequency. In a paper published in 2017 we proposed an expression for the dynamic shear compliance $J(\omega)\equiv 1/G(\omega)$, which was shown to fit well to squalane data over a sizable range of temperatures and frequencies \cite{hec17a}. The model also describes well data for 1-propanol \cite{wei21}. 

Squalane, which is a reference liquid for rheological measurements \cite{com13,sch15b}, has a strong beta process. The model proposed in Ref. \onlinecite{hec17a} was based on two assumptions: 1) The dynamic shear compliance is additive in the alpha and beta processes; 2) for $\omega\to\infty$ the alpha process obeys for its imaginary part $J''(\omega)\propto\omega^{-1/2}$ or, equivalently, $G''(\omega)\propto\omega^{-1/2}$. The second property can be justified by different arguments, many of which date far back in time \cite{gla60,isa66,BEL,dor70,mon70,maj71,kim78,lam78,wyl79,cun88,con89,lis97,dyr06a,dyr07}; this so-called $\sqrt{t}$-relaxation recently received strong support from light-scattering measurements on 16 supercooled liquids of quite different chemistry \cite{pab21}. An $\omega^{-1/2}$ high-frequency decay of a dynamic response function corresponds via the fluctuation-dissipation theorem to a short-time behavior of the relevant time-autocorrelation function that deviates from a constant by a term proportional to the square root of the time, hence the name $\sqrt{t}$-relaxation. Dielectric data also indicate that this short-time/high-frequency behavior of the alpha process may be universal \cite{nie09}, but the experimental evidence is here less convincing than for light scattering.

The model proposed in Ref. \onlinecite{hec17a} was based on an electrical equivalent circuit, which is reproduced in the Appendix where its translation into a rheological circuit is given. It leads to the following expression for the dynamic shear compliance in which $\tau_\alpha$ and $\tau_\beta$ are the characteristic times for the alpha and beta processes and $J_\alpha$ is the high-frequency limit of the compliance (often denoted by $J_\infty$)

\be\label{NBO}
J(\omega) = J_\alpha\left( 1 + \frac{1}{i\omega \tau_\alpha} +
\frac{k_1}{1+k_2\sqrt{i\omega\tau_\alpha}}\right) +
\frac{J_\beta}{1 + (i\omega\tau_\beta)^b}\,.
\ee
The first term is the contribution from the alpha process, which is a sum of a constant, a dc flow term, and a Cole-Cole type term \cite{col41} with exponent 1/2. The second term is the contribution from the beta process with compliance strength $J_\beta$, which is a Cole-Cole type term with exponent $b$. \Eq{NBO} has a total of seven parameters; the alpha term has four parameters: two of dimension ($J_\alpha$ and $\tau_\alpha$) and the two dimensionless shape parameters $k_1$ and $k_2$.

\begin{figure}
	\includegraphics[height=12cm]{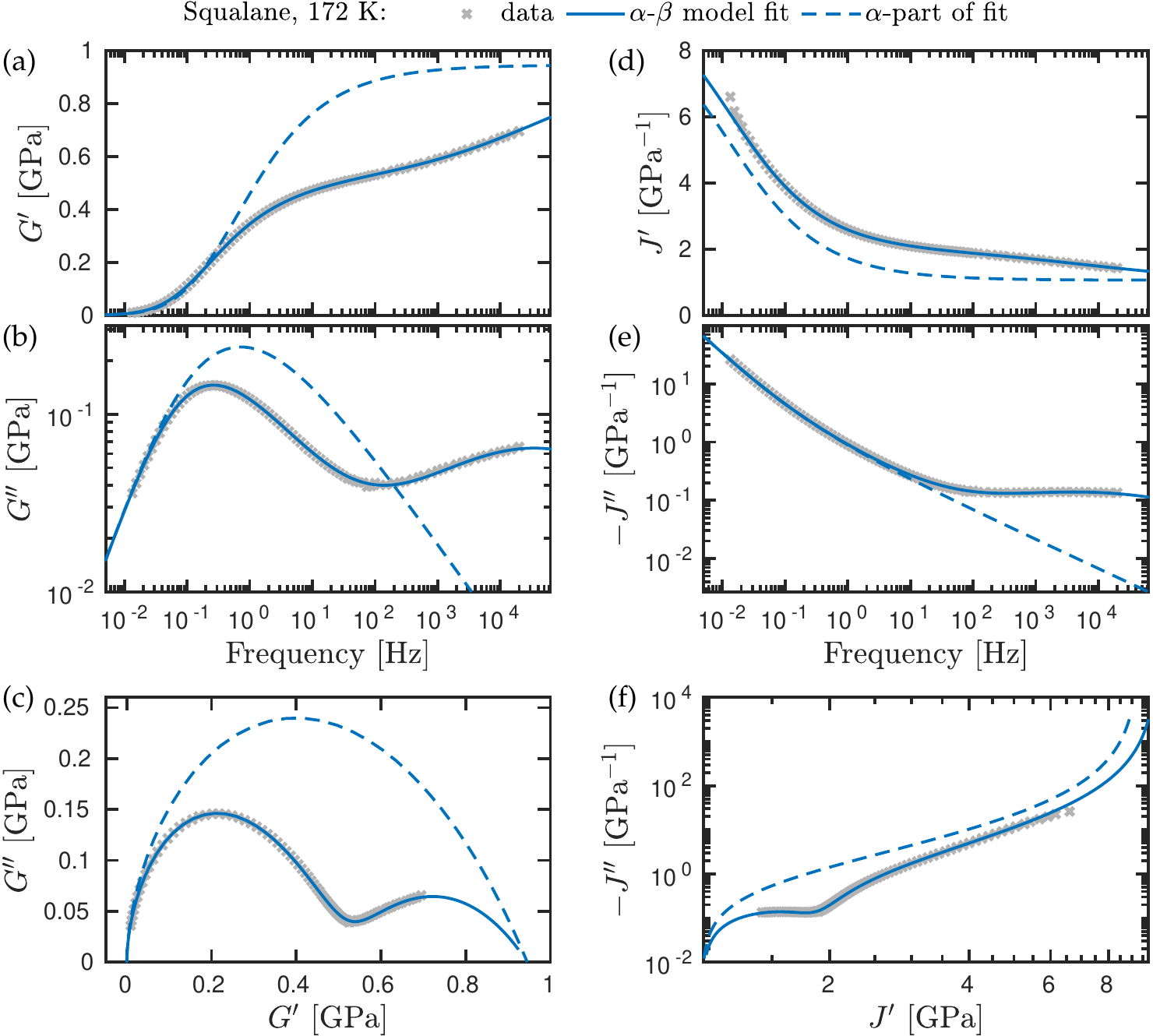}
	\caption{\label{fig1}Squalane data at 172 K (gray crosses) and the prediction of \eq{NBO} (full curves) in different representations (data and fits are from Ref. \onlinecite{hec17a}). (a) and (b) give the real and imaginary parts of the shear modulus, respectively, while (c) is the corresponding Nyquist plot. Figures (d), (e), and (f) give the same data in the compliance representation. In all the figures, the dashed curves are the predictions of the alpha part of \eq{NBO}, which does not provide a convincing fit to the low-frequency part of the data. In order to validate the alpha part of \eq{NBO} one needs to study liquids with little or no beta relaxations, which motivates the present paper.}
\end{figure}

An obvious question is the following: Is the excellent fit of \eq{NBO} to squalane data \cite{hec17a} due to the many fit parameters or does it reflect physical reality? If the latter is the case, the model provides a good starting point for arriving at a better understanding of the physics of supercooled liquids, for instance described in terms of independent and additive single-particle mean-square displacements deriving from the alpha and beta processes \cite{hec17a}. That would imply additivity of the alpha and beta response to an external field coupling to the single-particle motion, which by a Stokes-Einstein type argument translates into additivity of the shear compliance contributions from the alpha and the beta processes. Whether or not such an additivity of compliance reflects an element of physical reality cannot be judged from the squalane data of Ref. \onlinecite{hec17a}, however, because of the sizable beta process. 

To illustrate this we plot in \fig{fig1} squalane data for the dynamic shear modulus at the temperature $T=172$K. (a) and (b) show the real and imaginary parts as functions of the frequency ($\omega/2\pi$) while (c) shows a Nyquist plot, i.e., with the real part along the x-axis and the imaginary part along the y-axis. The data, which in (d), (e), and (f) are given in the compliance representation, are marked as gray crosses while the full curves are the best-fit model predictions (with the same parameters for both representations). The dashed curves are the predictions of the alpha part of \eq{NBO}. This analysis does not confirm that the alpha process of \eq{NBO} has independent physical reality. In order to properly investigate whether the first term of \eq{NBO} corresponds to an element of physical reality, one should compare this term to data for liquids with little or no mechanical beta relaxations. Although such liquids are rare, they do exist. This paper presents new data for the two silicone diffusion pump oils DC704 (tetramethyltetraphenyl trisiloxane) and DC705 (trimethylpentaphenyl trisiloxane). These liquids have no observable beta relaxation in the shear mechanics when probed with our PSG shear transducer covering more than six decades of frequency \cite{chr95}, and they are therefore ideal for testing the alpha part of \eq{NBO}. We show below that this ``alpha model'' fits well to the data and conclude that the alpha-beta  additivity of the dynamic compliance of \eq{NBO} is a physically meaningful assumption.

\begin{figure*}
	\includegraphics[width=\textwidth]{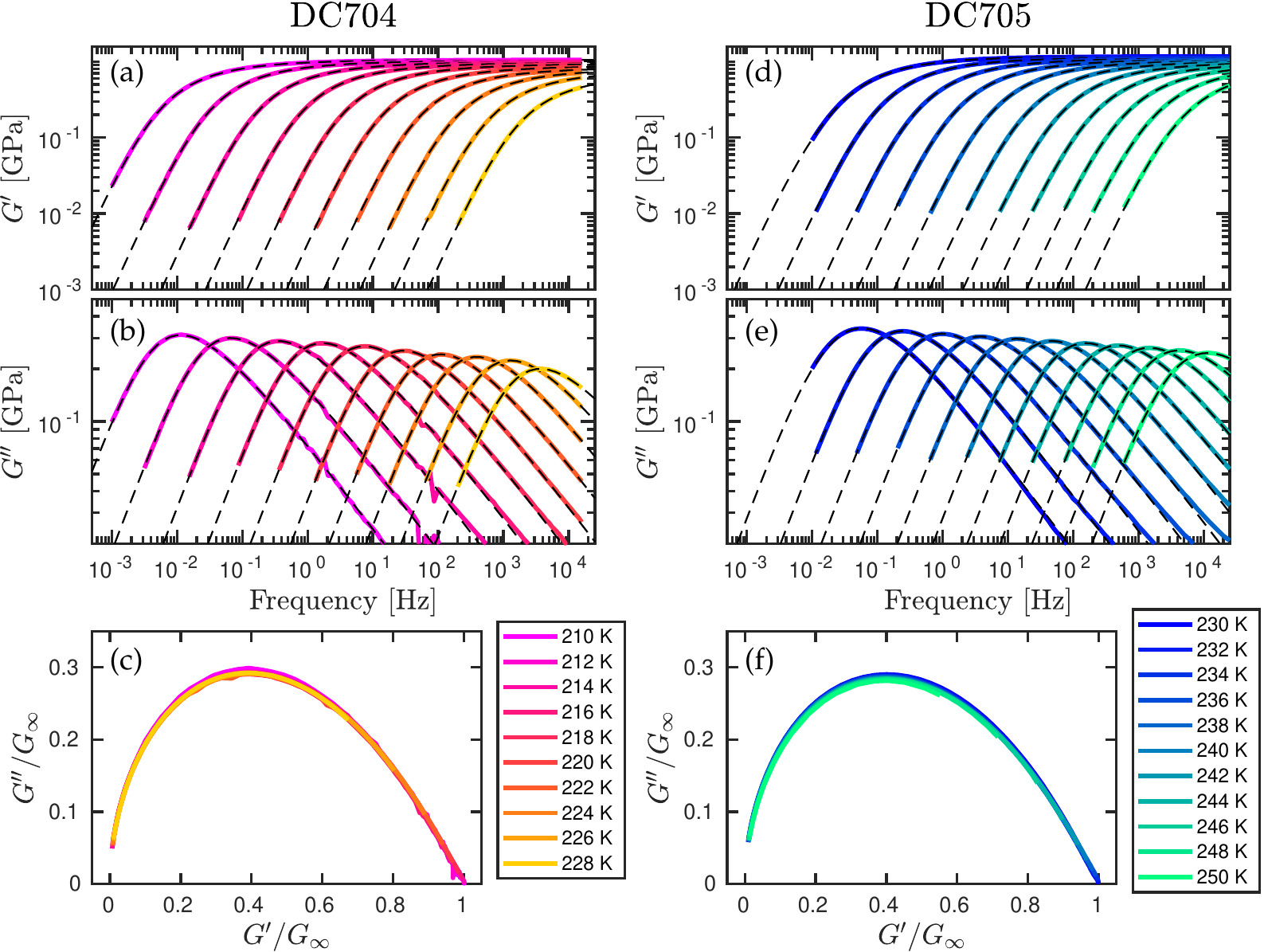}
	\caption{\label{fig_data}Dynamic shear modulus data for DC704 and DC705 over a range of temperatures. (a), (b), (d), and (e) give the real and imaginary parts of the frequency-dependent shear modulus $G(\omega)$ of the two liquids, while (c) and (f) present the data in normalized Nyquist plots. The dashed curves are best fits to \eq{NBO2}. }
\end{figure*}

\section{Data and model fits}\label{sec2}

Data for the dynamic shear-mechanical properties of DC704 and DC705 were obtained by the three-disc piezo-ceramic shear transducer described in Refs. \onlinecite{chr95} and \onlinecite{hec13}; references \onlinecite{iga08a} and  \onlinecite{iga08b} give details about the cryostat and the impedance-measuring setup. Data were obtained with 2~K intervals using a home-built cryostat that keeps the temperature stable within 10 mK. The data are from previously unpublished measurements but consistent with published data \cite{hec13}.

\Fig{fig_data} shows the dynamic shear modulus data as colored symbols in which the color reflects the temperature. The four upper figures show the real and negative imaginary parts of $G(\omega)=G'(\omega)+iG''(\omega)$, while (c) and (f) show the corresponding normalized Nyquist plots. The latter are sensitive to changes in the relaxation-time distribution, i.e., to violations of time-temperature superposition (TTS). While there is not exact TTS for either liquid, the deviations are minor. We observe a slight narrowing of the loss as the temperature is lowered, but it should be emphasized that this happens while at the same time the average relaxation time changes by more than five decades. In \fig{fig_data} the dashed curves are fits by the alpha part of \eq{NBO}, i.e., by $G(\omega)=1/J(\omega)$ in which

\be\label{NBO2}
J(\omega) = J_\alpha\left( 1 + \frac{1}{i\omega \tau_\alpha} + \frac{k_1}{1+k_2 \sqrt{i\omega\tau_\alpha}}\right)\,.
\ee
The rheological circuit corresponding to \eq{NBO2} is given in the Appendix.

\begin{figure*}
	\includegraphics[width=\textwidth]{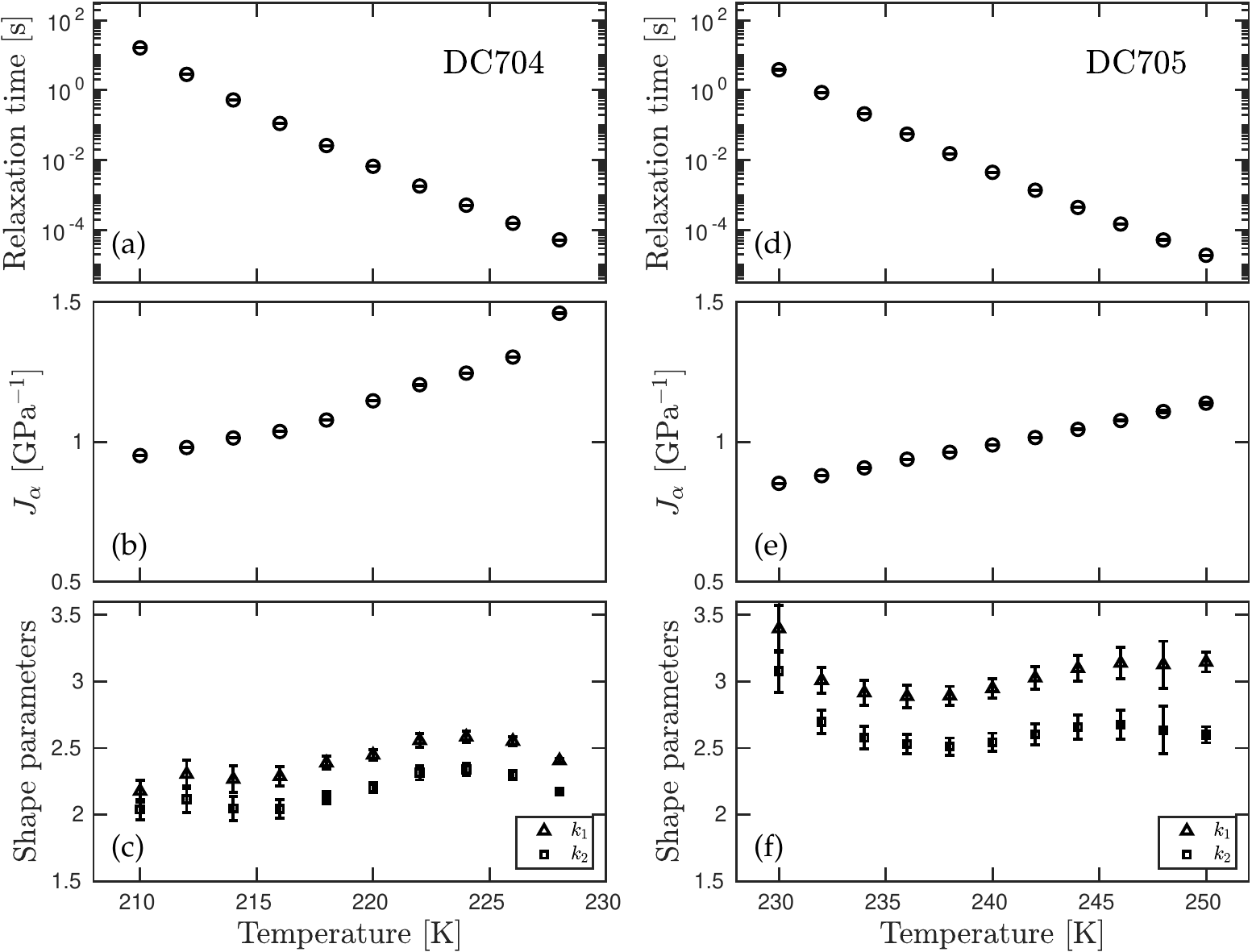}
	\caption{\label{fit_parametre}Best-fit parameters as functions of the temperature for DC704 and DC705 in which the vertical lines are error bars. (a) and (d) show the relaxation time $\tau_\alpha$, (b) and (e) show $J_\alpha$, while (c) and (f) show the two shape parameters $k_1$ and $k_2$. The fact that the latter are not temperature independent shows that time-temperature superposition (TTS) is not rigorously obeyed.} 
\end{figure*}

In the low-frequency limit of \eq{NBO2} one has $J(\omega)\cong J_\alpha/ ({i\omega \tau_\alpha})$. Since the dynamic shear viscosity $\eta(\omega)$ is calculated from the shear compliance by means of $\eta(\omega)=1/[i\omega J(\omega)]$ \cite{har76}, the dc viscosity $\eta_0$ is given by

\be\label{eta_expr}
\eta_0
\,=\,\frac{\tau_\alpha}{J_\alpha}\,.
\ee
In the high-frequency limit, \eq{NBO2} implies $J(\omega)\cong J_\alpha$ which means that the high-frequency plateau modulus $G_\infty$ is given by

\be\label{Ginf_expr}
G_\infty
\,=\,\frac{1}{J_\alpha}\,.
\ee
Combining \eq{eta_expr} with \eq{Ginf_expr} leads to the well-known Maxwell relation $\tau_\alpha=\eta_0/G_\infty$ \cite{har76}. 

\Fig{fit_parametre} shows the best-fit parameters as functions of the temperature. The increase of $J_\alpha$ with temperature is consistent with the well-known fact that $G_\infty$ usually decreases as the temperature is increased \cite{dyr96,dyr06}. The shape parameters are not entirely constant, confirming the already mentioned fact that TTS is slightly violated.

\begin{figure*}
	\includegraphics[width=\textwidth]{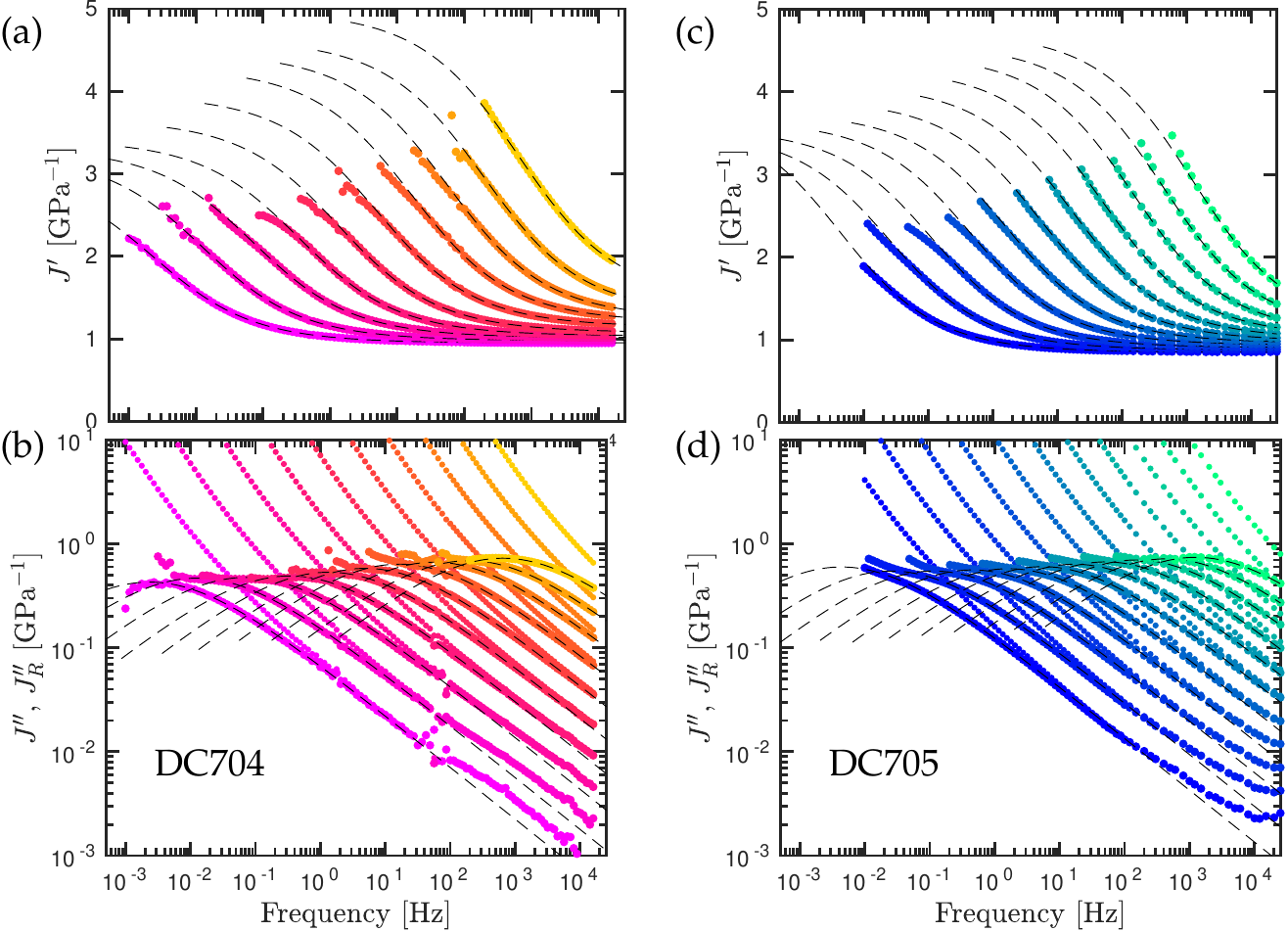}
	\caption{\label{J-fit}Real and imaginary parts of the shear compliance as functions of the frequency. The colored symbols are data, the black dashed lines are model fits. The data do not confirm the alpha-model predction of a low-frequency plateau of the real part ((a) and (c)), but there are indications of the corresponding peak in the imaginary part of the recoverable shear compliance defined in \eq{JR} (lower curves of (b) and (d)). }
\end{figure*}

The dynamic shear modulus has a different emphasis than the dynamic shear compliance $J(\omega)=J'(\omega)-iJ''(\omega)$, and it is important to investigate whether the model also fits data well in the latter representation. \Fig{J-fit} shows the real and imaginary parts of $J(\omega)$ where the colored symbols are the data and the black dashed lines are the model fits (with the same parameters as in \fig{fig_data}, compare \fig{fit_parametre}). The model predicts a low-frequency plateau for $J'(\omega)$, compare \eq{NBO}, which is not possible to measure with our setup that is optimized for moduli above $10^{7}$ Pa. The data do not show evidence for the plateau prediction, but they also do not contradict it. If the purely imaginary flow term $1/(\eta_0i\omega)$ is subtracted from $J(\omega)$ to form the ``recoverable shear compliance'' \cite{pla65,buc16,buc17}, which except for a constant is identical to the so-called retardational compliance \cite{har76}, i.e.,

\be\label{JR}
J_R(\omega)
\,\equiv\, J(\omega)-J_\alpha/(i\omega\tau_\alpha)=J(\omega)-1/(\eta_0i\omega)\,,
\ee
there are hints of a peak in the imaginary part as predicted by the model (\fig{J-fit}(b) and (d)).

\begin{figure*}
  \includegraphics[width=\textwidth]{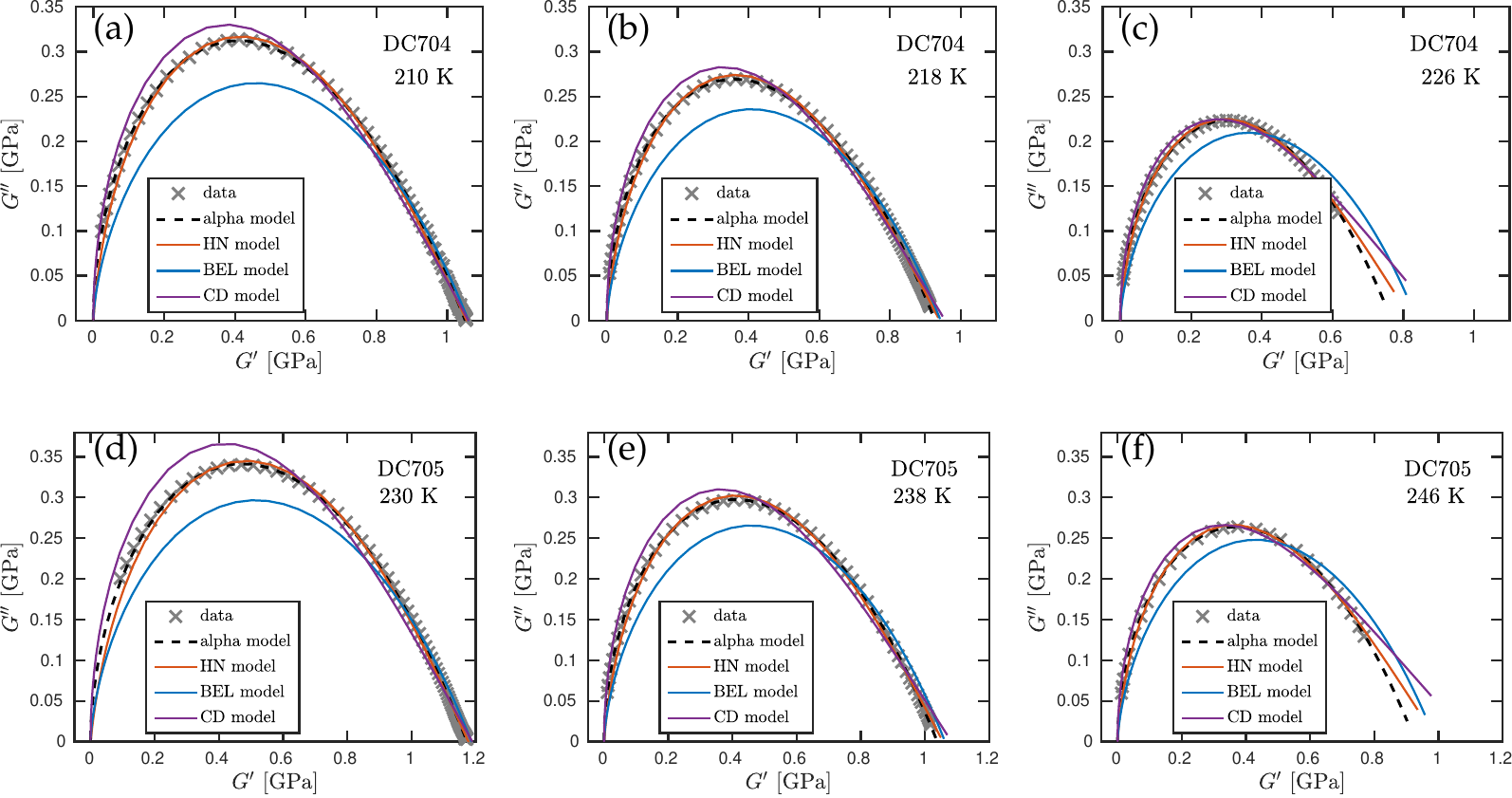}
\caption{\label{Nyquist}Nyquist plots of the $G(\omega)$ data at selected temperatures compared to the best fits of different models. The fits of the alpha and the HN models are all good, whereas the fits by the BEL and CD models deviate notably from the data.}
\end{figure*}

\section{Comparing to fits by three other models}\label{sec3}

This section compares to fits by other models. Recalling that $G(\omega)=1/J(\omega)$, it is sometimes convenient to formulate a model in terms of one or the other representation, e.g., writing the dynamic shear modulus as $G_\infty$ minus a function commonly used for fitting dielectric relaxation data, as done in two cases below. For all three functions, the dynamic shear-modulus goes to zero at low frequencies and reaches a plateau at high frequencies as required for any realistic model of an alpha process. 

\begin{itemize}

\item The Havriliak-Negami (HN) type model is defined \cite{hec17a} by

\be\label{HN}
G(\omega)
\,=\, \frac{1}{J_\alpha}\,\left( 1 - \frac{1}{(1+(i\omega  \tau_\alpha)^c)^a}\right)\,.
\ee

\item The Barlow-Erginsav-Lamb (BEL) model is defined \cite{BEL} by 

\be\label{BEL}
J(\omega)
\,=\,\left(\frac{1}{\sqrt{G_\infty}}+\frac{1}{\sqrt{i\omega\eta_0}}\right)^2
\,=\, \frac{1}{G_\infty}+\frac{1}{i\omega\eta_0}+\frac{2}{\sqrt{ G_\infty i\omega\eta_0}}\,.
\ee
This can be obtained from the alpha model \eq{NBO2} by substituting $k_1=2$ and $k_2=1$ and replacing unity by zero in the last denominator. The latter corresponds to removing a spring in the rheological circuit, compare the Appendix, thus arriving at

\be\label{BEL2}
J(\omega) = J_\alpha\left( 1 + \frac{1}{i\omega \tau_\alpha} + \frac{2}{ \sqrt{i\omega\tau_\alpha}}\right)\,.
\ee
From this perspective, \eq{NBO2} may be regarded as an extension of the BEL model by introducing two shape parameters.

\item The Cole-Davidson (CD) type model is defined by

\be\label{CD}
G(\omega)
\,=\, \frac{1}{J_\alpha}\,\left( 1 - \frac{1}{({1+i\omega \tau_\alpha})^a}\right)\,.
\ee

\end{itemize}
In all cases $J_\alpha$ is the high-frequency limit of $J(\omega)$ and $J_\alpha=1/G_\infty$.

\begin{figure*}
  \includegraphics[width=\textwidth]{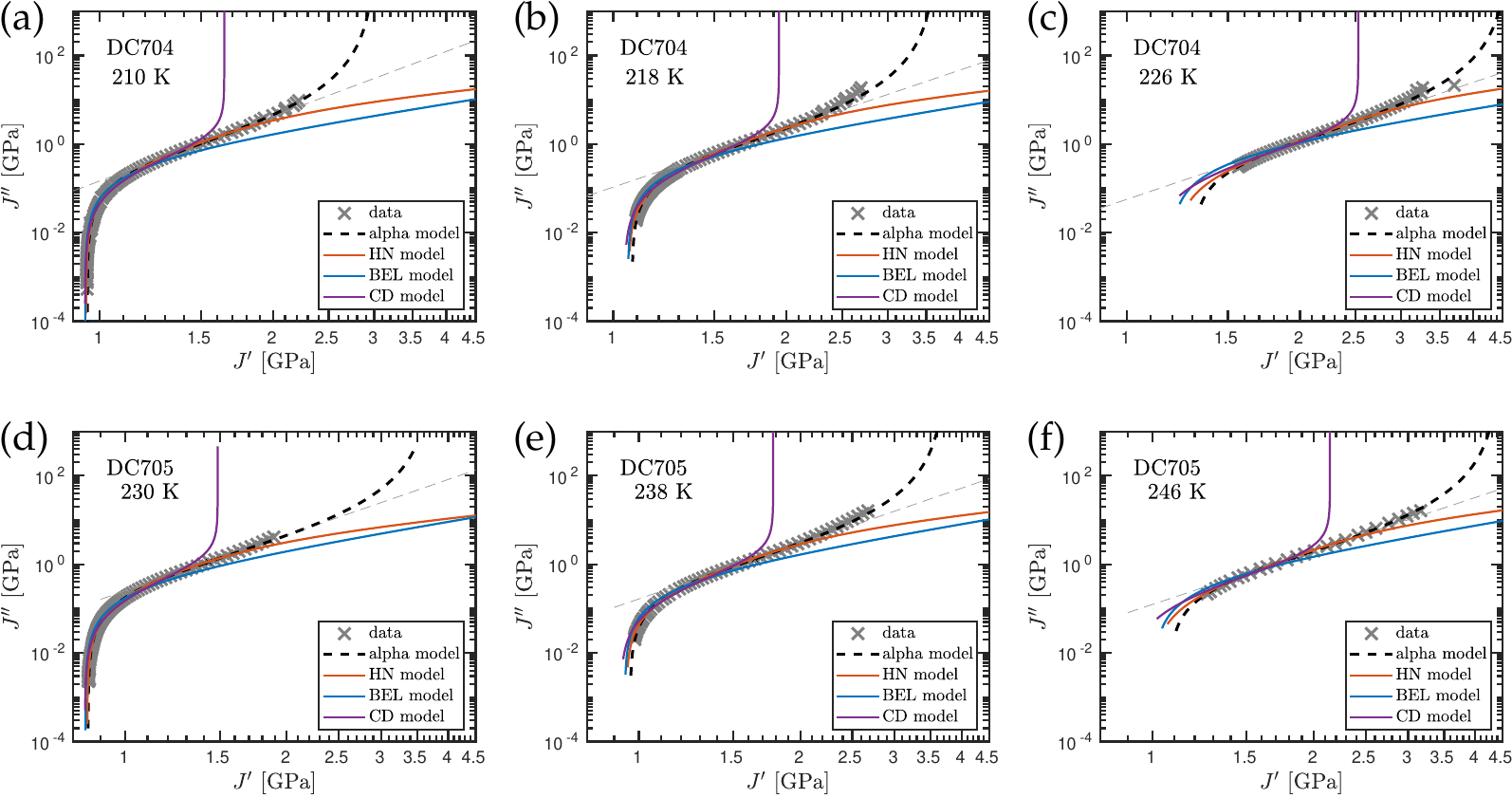}
\caption{\label{LogNyquist}Logarithmic Nyquist plots of compliance data and fits at selected temperatures. The approach to vertical tangents for the alpha and HN models reflects the existence of a low-frequency plateau in the real part of the shear compliance in these two models. The approach to a plateau is fitted best by the alpha model.}
\end{figure*}

All models have two parameters of dimension, one determining the scale of the time and one determining the scale of the compliance. In regard to the number of shape parameters, the alpha and HN models have two such parameters, the BEL model has none, and the CD model has one. This, of course, influences the quality of the fits obtained, compare \fig{Nyquist} that shows all data and fits in Nyquist plots of the dynamic shear modulus. We see  generally good fits of the alpha and HN models, while the BEL and and CD models do not give good fits. 

In order to better be able to compare the two best models according to \fig{Nyquist} -- the alpha and HN models -- we introduce in \fig{LogNyquist} a ``log-compliance'' Nyquist plot. This way of showing data has a different focus than the standard Nyquist plot of \fig{Nyquist}. In particular, one can here investigate the low-frequency real part of the compliance, which has a plateau for the alpha and CD models, but not for the BEL and HN models. \Fig{LogNyquist} shows that the approach to a plateau is best represented by the alpha model.

\begin{figure*}
    \centering
    \includegraphics[height=5cm]{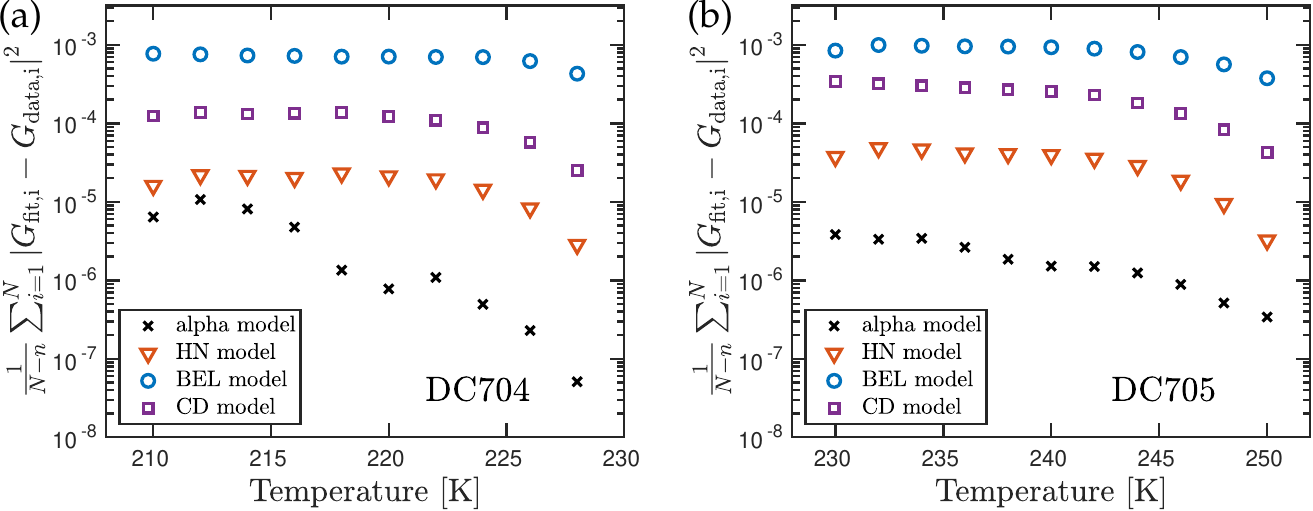}
    \caption{Quality of all fits at all temperatures measured as indicated by the squared difference between data and model prediction of the dynamic shear modulus averaged over all $N$ measured frequencies in which $n$ is the number of fit parameters that is 4 for the alpha and HN models, 3 for the CD model, and 2 for the BEL model. For both liquids, the alpha model is superior at all temperatures.   \label{fig:goodness}}
\end{figure*}

\Fig{fig:goodness} compares the quality of all fits by giving at each temperature the average deviation between data and best-fit predictions. For both liquids the alpha model is best at all temperatures.

\section{Summary}

Using shear-dynamical data for DC704 and DC705 this paper has investigated the question to which extent the separation of the compliance function proposed in Ref. \onlinecite{hec17a} into additive alpha and a beta parts makes sense physically. If this were the case, then the alpha part of the model should fit well to data for liquids with no mechanical beta relaxations. The two silicone oils studied here fall into this category.

The alpha model fits data well for both DC704 and DC705. We take this as a confirmation that the simple $\omega^{-1/2}$ high-frequency behavior of the imaginary parts of the modulus and compliance reflects an important characteristic of the two liquids studied, confirming previous and recent works \cite{dyr07,nie09,pab21}. An important further test of the alpha model would be to check whether the predicted low-frequency plateau in the real part of the shear compliance exists and is predicted correctly, because this feature distinguishes the alpha model from some of the other models studied here. Our shear transducer \cite{chr95} is optimized for high-modulus samples and cannot be used for reliable measurements at the relevant low frequencies, unfortunately. Such measurements are important also because they would allow for testing the prediction of the alpha model that the Nyquist plot of $J_R(\omega)$ (\eq{JR}) is symmetrical and identical to the standard dielectric relaxation Nyquist plot of the Cole-Cole function with exponent 1/2 \cite{col41}.

\begin{acknowledgments}
	This work was supported by the VILLUM Foundation's \textit{Matter} grant (16515).
\end{acknowledgments}

\section*{Data availability}

The data that support the findings of this study are available from the corresponding authors upon request.

\section*{Appendix: Diagrams}

This Appendix presents the rheological model behind \eq{NBO} in two equivalent formulations: the electrical-equivalent circuit used in Ref. \onlinecite{hec17a} and the spring-dashpot formulation that is standard in rheology. Even though not visually obvious, these two formulations are mathematically equivalent. In the present case of highly viscous glass-forming liquids where intertia plays no role, a parallel combination of electrical circuit elements corresponds to a mechanical series combination and \textit{vice versa}. This is because currents are additive for an electrical parallel combination and velocities (currents) are additive in a mechanical series combination. Likewise, the potential drop across an electrical series combination is additive and forces/stresses (potential drops) are additive for a mechanical parallel combination.

For translating between the two types of circuits, besides the above parallel/series interchange one reasons as follows. Electrical charge corresponds to (shear) displacement and electrical current corresponds to shear rate. A resistor in an electrical circuit is an element for which the current is proportional to the potential drop. This corresponds to a dashpot in a rheological circuit, the element characterized by a constant frequency-independent viscosity for which the shear rate is proportional to the force (stress). A capacitor in an electrical circuit is an element for which the charge is proportional to the potential drop, which translates into a spring in a rheological  circuit for which the (shear) displacement is proportional to the force (stress).

\begin{figure*}
    \centering
    \includegraphics[width=\textwidth]{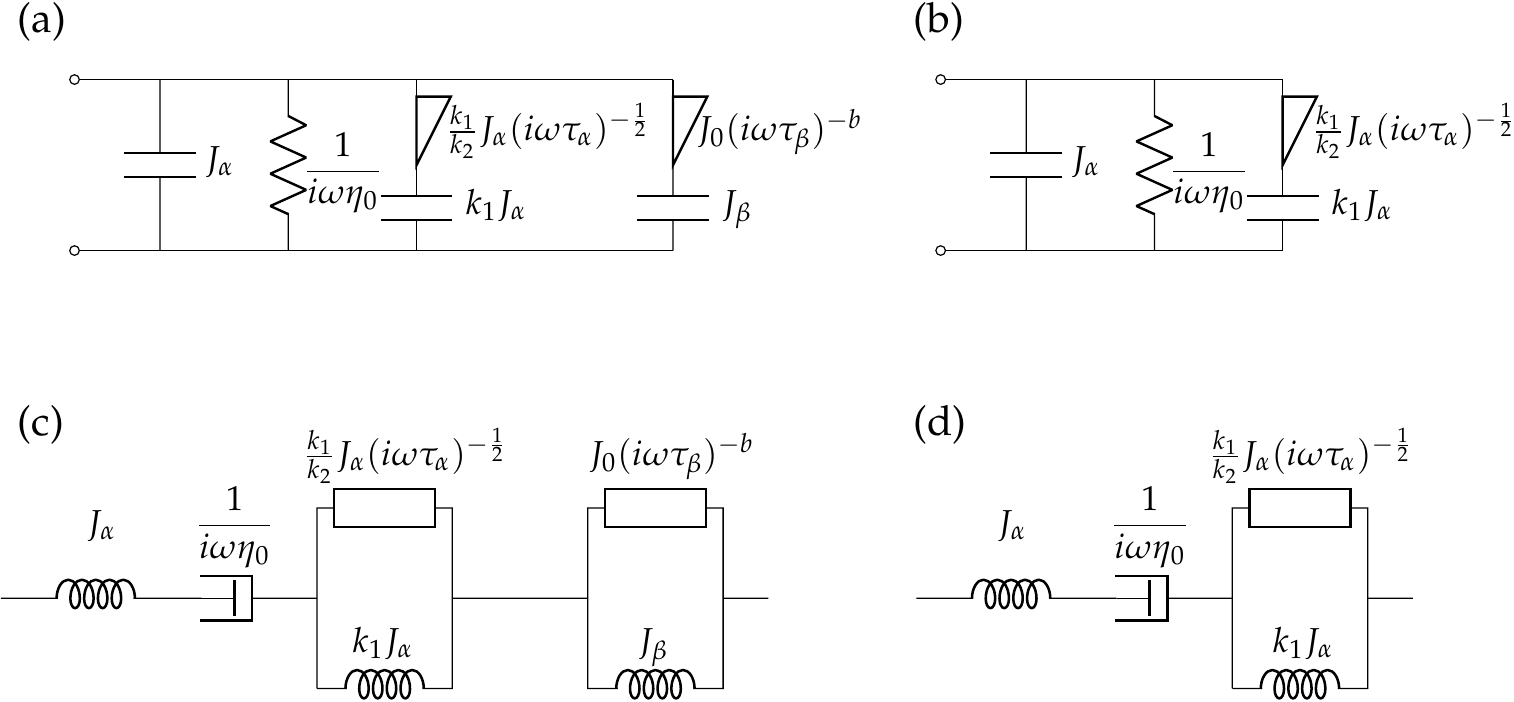}
    \caption{Diagrammatic representations of shear rheological models.
    (a) Electrical-equivalent diagram of the alpha-beta model of Ref. \onlinecite{hec17a}. A triangle is a constant-phase element characterized by an admittance that varies with frequency as $(i\omega)^{-n}$, where $n=1/2$ for the alpha element while $n=b$ is a free parameter for the beta element. 
    (b) The alpha part of the alpha-beta model, i.e., the electrical equivalent circuit of \eq{NBO}.
    (c) Rheological model corresponding to the alpha-beta electrical-equivalent diagrams. Besides the standard spring and dashpot symbols, a box is a constant-phase element with the indicated power-law frequency dependence of the shear compliance. 
    (d) Rheological model corresponding to the alpha model \eq{NBO}.
    \label{fig:diagrams}}
\end{figure*}

With this in mind, \fig{fig:diagrams} shows the relevant diagrams. \Fig{fig:diagrams}(a) reproduces the electrical equivalent circuits of Ref. \onlinecite{hec17a} and (b) is the alpha part of this circuit. Note that these electrical diagrams are \textit{equivalent diagrams} of the mechanical response; they are not diagrams for the dielectric response. In particular, no additivity is implied for the alpha and beta dielectric responses. The rheological diagram of the alpha model is given in (d), corresponding to the electrical equivalent circuit (b). To arrive at \eq{NBO} for the (shear) compliance one uses the fact that compliances are additive for a mechanical series connection (the force is the same for each element while displacements are added).

\end{document}